\documentclass{article}%
\usepackage{amsmath}
\usepackage{graphicx}%
\usepackage{amsfonts}%
\usepackage{amssymb}
\newtheorem{theorem}{Theorem}

\newtheorem{proposition}[theorem]{Proposition}
\newtheorem{remark}[theorem]{Remark}

\newenvironment{proof}[1][Proof]{\textbf{#1.} }{\ \rule{0.5em}{0.5em}}

\def\Tr{\mathop\mathrm{Tr}\nolimits}

\begin{document}

\title{The quantum measurement problem enhanced}
\author{Gebhard Gr\"{u}bl\\Institut f\"{u}r Theoretische Physik der 
Universit\"{a}t Innsbruck,\\Technikerstr. 25, A-6020 Innsbruck,
Austria\\[5pt]E-mail: gebhard.gruebl@uibk.ac.at}
\date{}
\maketitle

\begin{abstract}
The quantum measurement problem as formalised by Bassi and Ghirardi [Phys.
Lett. A 275 (2000)373] without taking recourse to sharp apparatus observables
is extended to cover impure initial states.
\end{abstract} PACS: 03.65.Bz

\section{Introduction}

In a recent letter \cite{BG} Bassi and Ghirardi have developed a strong and
still very simple formulation of the quantum measurement problem
(QMP)\footnote{melodramatised by Schr\"{o}dinger's cat} within standard
quantum theory. They derived the nonoccurrence of definite apparatus
configurations from assumptions considerable weaker than the ones of von
Neumann's original QMP \cite{Neu}. They carefully justified their assumptions
from a standard quantum theoretical perspective and removed certain
oversimplifications from von Neumann's treatment. The latter had been used by
some authors (quoted in \cite{BG}) as loopholes to deny the QMP's very existence.

Bassi and Ghirardi take the premeasurement state as a tensor product of two
pure states. One for the microsystem to be measured and one for the
environment incorporating the apparatus. If each individual quantum system
indeed possesses\ a pure state, then the Bassi Ghirardi QMP is sufficiently
strong, since if individual systems do not develop definite apparatus
configurations, an ensemble of different pure states cannot do so either. Yet
standard quantum theory suggests that individual systems in general have
impure states. The argument is as follows. If a composite system with Hilbert
space $\mathcal{H}_{1}\otimes\mathcal{H}_{2}$ has a pure state density
operator $\rho$, its subsystem $1$ has the induced density operator $\rho_{1}$
defined through: $\Tr\left(  \rho\left(  A\otimes id_{2}\right)  \right)
=\Tr(\rho_{1}A)$ for all linear continuous $A:\mathcal{H}_{1}\rightarrow
\mathcal{H}_{1}$. In general $\rho_{1}$ is impure. This fact leads to the idea
that impure states should be considered as equally fundamental as pure ones.
They should not be associated with ensembles only, but with individual systems
too. Therefore from the perspective of standard quantum theory a weakness of
the Bassi Ghirardi QMP lies in its limitation to pure states and it seems
necessary to investigate whether the argument of Bassi and Ghirardi can be
generalised to impure premeasurement states.

A sequence of successively more general formulations of the QMP for impure
states already exists. These works by Wigner \cite{Wig}, d'Espagnat
\cite{dEs}, Fine \cite{Fin}, Shimony \cite{Shi}, Busch and Shimony \cite{BS},
Busch \cite{B}, they all derive from varying assumptions that the
postmeasurement state $\rho$ does not equal a convex sum $\sum_{i}p_{i}%
\rho_{i}$ (i.e. $\sum_{i}p_{i}=1,\ p_{i}>0$) of mutually exclusive states
$\rho_{i}$ with definite apparatus configurations. (Here the apparatus
configuration of $\rho_{i}$ is assumed different from the one of $\rho_{j}$
for $i\neq j$.) Since the apparatus configuration is definite in the trivial
case of a one term sum, i.e. $\rho=\rho_{i}$, only, the postmeasurement
apparatus configuration is indefinite. Note that even if $\rho$ would equal a
convex sum of the above type with at least two terms, the apparatus
configuration still were indefinite - at least if impure states are associated
with individual systems. Therefore the nonexistence of the above convex
representation proves much more than is needed for having a measurement
problem, because for the QMP it is completely irrelevant whether the
postmeasurement state is diagonal with respect to the apparatus observable or
not \cite{Bell}.\footnote{The diagonalisability obtains relevance if states
are associated with ensembles exclusively. Then a convex decomposition of a
state is understood as a mixture of other ensembles. Such a strict ensemble
interpretation, however, is unable to associate a state with an ensemble if
the ensemble is understood as a single composite system. Or you may ask how
many members needs an ensemble to be an ensemble.} The QMP arises whenever the
apparatus observable has nonzero variance.

In all the above impure state formulations of the QMP different definite
apparatus configurations are associated with mutually orthogonal subspaces of
the problem's Hilbert space. It is assumed that the ranges of two density
operators $\rho_{1}$ and $\rho_{2}$ with different apparatus configurations
are orthogonal to each other. Yet it is exactly this conception which has been
abandoned by Bassi and Ghirardi for pure states. Therefore the question still
stands whether the argument of Bassi and Ghirardi can be generalised to impure
premeasurement states.

I shall present here a QMP which also covers the case of impure states but
avoids associating different apparatus configurations with orthogonal
subspaces. This realises what is called unsharp pointer reading or unsharp
objectification in \cite{B}.

The debate on whether the decoherence program has resolved the QMP continues
without emerging consensus \cite{Adl}. Clearly, any proposed resolution of a
specific QMP should, as a first consistency check, exhibit which of the QMP's
assumptions are violated, in order to circumvent its conclusion. Bohmian
mechanics for instance harshly violates the assumption 2 of section 3 and
therfore is able to resolve the QMP in a surprisingly simple and eye-opening
way \cite{BDDGZ}, \cite{DGZ}, \cite{Due}. The apparatus configuration is
determined by Bohmian positions and has little to do with the wave function
part of the state. Almost all of the above discussion is absurd from a Bohmian
point of view. Another proposal to resolve the QMP violates assumption 3 of
section 3: stochastic state reduction models give up the unitary time
evolution \cite{Adle}. Since the decoherence program employs the minimal
quantum structures only it necessarily seems incapable to resolve the QMP.

In order to better display the weakening of assumptions, I shall distill in
section 2 the findings of \cite{BG} into a tight proposition. Here I shall not
duplicate the ample discussion, which Bassi and Ghirardi gave in order to
support their assumptions. The proof of their QMP, however, will be redone. In
section 3 I shall state and prove the stronger QMP covering impure states.
This proposition contains the one of section 2 as a special case. Finally I
speculate on the impact which the QMP might bring to the future developement
of quantum theory.

\section{QMP for pure states}

Bassi and Ghirardi made the following assumptions. Let $\mathcal{H}_{s}$ be
the microsystem's Hilbert space from which two vectors $\psi_{1},\psi_{2}$ are
chosen such that $\left\langle \psi_{i},\psi_{j}\right\rangle =\delta_{i,j}$.
The environment (including the measurement apparatus and the observer) has the
Hilbert space $\mathcal{H}_{E}$. The unit sphere in the total space
$\mathcal{H}_{s}\otimes\mathcal{H}_{E}$ is assumed to contain subsets
$A_{1},A_{2}$ such that the observer perceives the apparatus in the distinct
configuration $i$ if and only if the total system's pure state is represented
by a vector belonging to $A_{i\text{.}}$ The sets $A_{i}$ are supposed to be
separated from each other in the following sense. For some fixed positive
value $\varepsilon<1/2$ the inequality
\begin{equation}
\left|  \left\langle x,y\right\rangle \right|  ^{2}<\varepsilon
\label{distpure}%
\end{equation}
holds for all $x\in A_{1}$ and for all $y\in A_{2}$. The initial state unit
vectors are assumed to be given by $\psi\otimes e$, where $e$ belongs to the
set
\[
P=\left\{  e\in\mathcal{H}_{E}\mid\left\|  e\right\|  =1\text{ and }%
\psi\otimes e\in A_{1}\text{ for all unit vectors }\psi\in\mathcal{H}%
_{s}\right\}  .
\]
Thus initially the observer perceives apparatus configuration $1$. The
measurement interaction results in a unitary operator $U$ on $\mathcal{H}%
_{s}\otimes\mathcal{H}_{E}$ giving the time evolution of states to an instant
of time at which this interaction has come to an end. Definite apparatus
configurations are supposed to result from $U$ for the initial states
$\psi_{i}\otimes e$, i.e. $U\left(  \psi_{i}\otimes e\right)  \in A_{i}$ is
assumed to hold for all $e\in P$ and for $i=1,2$.

\begin{proposition}
The vector $\left(  \psi_{1}+\psi_{2}\right)  /\sqrt{2}$\ has a neighborhood
$\mathcal{D}$ in $\mathcal{H}_{s}$ such that for all unit vectors $\psi
\in\mathcal{D}$ and for all $e\in P$ holds $U\left(  \psi\otimes e\right)
\notin A_{1}\cup A_{2}$.
\end{proposition}

\begin{remark}
An initial state $\psi\otimes e$ with apparatus configuration $1$ and $\psi$
sufficiently close to $\frac{1}{\sqrt{2}}\left(  \psi_{1}+\psi_{2}\right)
$\ does not develop a definite postmeasurement configuration (either $1$ or
$2$) under $U$.
\end{remark}

\begin{proof}%
\begin{gather*}
\left|  \left\langle U\left(  \frac{1}{\sqrt{2}}\left(  \psi_{1}+\psi
_{2}\right)  \otimes e\right)  ,U\left(  \psi_{i}\otimes e\right)
\right\rangle \right|  ^{2}=\\
=\frac{1}{2}\left|  \left\langle \left(  \psi_{1}+\psi_{2}\right)  \otimes
e,\psi_{i}\otimes e\right\rangle \right|  ^{2}=\frac{1}{2}\left|  \left\langle
\psi_{i},\psi_{i}\right\rangle \right|  ^{2}\left|  \left\langle
e,e\right\rangle \right|  ^{2}=\frac{1}{2}>\varepsilon.
\end{gather*}
Since $\psi\mapsto\left\langle \psi,\psi_{i}\right\rangle $ is continuous,
$\left|  \left\langle U\left(  \psi\otimes e\right)  ,U\left(  \psi_{i}\otimes
e\right)  \right\rangle \right|  ^{2}>\varepsilon$ for all $\psi$ in a
neighborhood $\mathcal{D}$ of $\frac{1}{\sqrt{2}}\left(  \psi_{1}+\psi
_{2}\right)  $ in $\mathcal{H}_{s}$. Because of $U\left(  \psi_{1}\otimes
e\right)  \in A_{1}$, the vector $U\left(  \psi\otimes e\right)  $ for any
$\psi\in\mathcal{D}$ is closer to $A_{1}$ than any vector from $A_{2}$.
Therefore $U\left(  \psi\otimes e\right)  \notin A_{2}$. Similarly one obtains
$U\left(  \psi\otimes e\right)  \notin A_{1}$ for all $\psi\in\mathcal{D}$.
\end{proof}

\section{QMP for impure states}

As in the former section the total system is divided into a microsystem and an
environment. The environment comprises the observer and the measurement
apparatus. Since the environment may be entangled with rest of the world,
which is ignored in the measurement interaction, it may not have a pure state
even at the level of individual systems. According to standard quantum theory
its state is generally given by an induced density operator, which is obtained
by restricting the expectation value functional to those observables which are
sensitive to the environment's degrees of freedom only. Imperfections in the
preparation of the microsystem may result in impure states for individual
microsystems as well. Thus one has to address initial states which factorise
into two density operators. Density operators are Hilbert Schmidt operators. I
shall make use of this fact and therefore I recall some properties of Hilbert
Schmidt operators \cite{RS} now.

A linear continuous operator $a$ on a separable Hilbert space $\mathcal{H}$
with scalar product $\left\langle \cdot,\cdot\right\rangle $\ belongs to the
set $\mathcal{C}_{HS}(\mathcal{H})$ of Hilbert Schmidt operators if and only
if $\Tr(a^{\ast}a)<\infty$. The set $\mathcal{C}_{HS}(\mathcal{H})$ is a
complex vector space with respect to the addition of operators. The Hilbert
Schmidt scalar product of $\mathcal{C}_{HS}(\mathcal{H})$ is defined as
$\left\langle a,b\right\rangle _{HS}:=\Tr(a^{\ast}b)$. The associated Hilbert
Schmidt norm is
\[
\left\|  a\right\|  _{HS}:=\sqrt{\Tr(a^{\ast}a)}.
\]
$\mathcal{C}_{HS}(\mathcal{H})$ is complete with respect to the norm
$\left\|  \cdot\right\|  _{HS}$. Thus $\left\langle \cdot,\cdot\right\rangle
_{HS}$ makes $\mathcal{C}_{HS}(\mathcal{H})$ into a Hilbert space. For all
$a\in\mathcal{C}_{HS}(\mathcal{H})$ and for all $b\in\mathcal{C}%
(\mathcal{H})$ (the algebra of linear continuous operators on $\mathcal{H}%
$)\ there holds $ab\in\mathcal{C}_{HS}(\mathcal{H})$ and $ba\in
\mathcal{C}_{HS}(\mathcal{H})$. Thus $\mathcal{C}_{HS}(\mathcal{H})$ is a
left/right ideal of $\mathcal{C}(\mathcal{H})$.

For $\psi\in\mathcal{H}\setminus0$ the orthogonal projection onto
$\mathbb{C}\cdot\psi$ is denoted as $P_{\psi}$. It holds
\begin{gather*}
\left\langle P_{\psi_{1}},P_{\psi_{2}}\right\rangle _{HS}=\frac{1}{\left\|
\psi_{1}\right\|  ^{2}.\left\|  \psi_{2}\right\|  ^{2}}\Tr\left(  \psi
_{1}\left\langle \psi_{1},\psi_{2}\right\rangle \left\langle \psi_{2}%
,\cdot\right\rangle \right) \\
=\frac{\left|  \left\langle \psi_{1},\psi_{2}\right\rangle \right|  ^{2}%
}{\left\|  \psi_{1}\right\|  ^{2}.\left\|  \psi_{2}\right\|  ^{2}}\in\left[
0,1\right].
\end{gather*}

Thus $\left\|  P_{\psi}\right\|  _{HS}=1$. More generally, since $\left\|
\rho\right\|  _{HS}^{2}=\Tr(\rho^{\ast}\rho)=\Tr(\rho^{2})\leq1$, the set of
density operators on $\mathcal{H}$ is a subset of $\mathcal{C}_{HS}%
(\mathcal{H})$. The equality $\left\|  \rho\right\|  _{HS}=1$ holds if and
only if $\rho^{2}=\rho$ (i.e. $\rho$ is a pure state). Observe that $\left\|
\sqrt{\rho}\right\|  _{HS}=1$ for any density operator, i.e. the square root
of a density operator belongs to the unit sphere in $\mathcal{C}%
_{HS}(\mathcal{H})$. Here $\sqrt{\rho}$ denotes that unique positive operator
which obeys $\left(  \sqrt{\rho}\right)  ^{2}=\rho$. For pure states holds
$\sqrt{\rho}=\rho$.\\
\\
I shall address the measurement problem now.

\begin{itemize}
\item Assumption 1: The microsystem has the (separable) Hilbert space
$\mathcal{H}_{s}$. Two vectors $\psi_{1},\psi_{2}\in H_{s}$ obeying
$\left\langle \psi_{i},\psi_{j}\right\rangle =\delta_{i,j}$ are chosen. The
environment has the (separable) Hilbert space $\mathcal{H}_{E}$. The total
Hilbert space is $\mathcal{H}:=\mathcal{H}_{s}\otimes\mathcal{H}_{E}$.

\item Assumption 2: The set of density operators on the total Hilbert space is
supposed to contain two subsets $A_{1},A_{2}$ with the property that the
apparatus is (perceived) in configuration $i=1,2$ if and only if the total
system's density operator $\rho$ belongs to $A_{i}$. The sets $A_{i}$ are
assumed to be separated from each other in the following sense, which also
makes them disjoint. There exists a positive real number $\varepsilon<1/2$
such that for all $\rho_{1}\in A_{1},\rho_{2}\in A_{2}$ the inequality
\begin{equation}
\left|  \left\langle \sqrt{\rho_{1}},\sqrt{\rho_{2}}\right\rangle
_{HS}\right|  <\varepsilon<\frac{1}{2} \label{distmixed}%
\end{equation}
holds. In case of pure states this condition specialises to the analogous
distance condition $\left(  \ref{distpure}\right)  $ of section 2.
\end{itemize}

\begin{remark}
Inequality $\left(  \ref{distmixed}\right)  $ is weaker than the usual
assumption that the apparatus configurations $i=1,2$ are associated with two
orthogonal projections $P_{i}$, projecting onto mutually orthogonal subspaces
of $\mathcal{H}$, eigenspaces of some apparatus observable. Thus $P_{i}%
P_{j}=\delta_{i,j}P_{i}$ holds. A density operator $\rho$ with apparatus
configuration $i$ is assumed to give the projection $P_{i}$ the expectation
value $1$. Thus $\Tr(\rho P_{i})=1$ holds for $\rho$ having apparatus
configuration $i$. Now $\Tr(\rho P_{i})=1$ is equivalent to $P_{i}\rho
P_{i}=\rho$. From this then follows for density operators $\rho_{i}$ with
apparatus configurations $i$ that
\begin{gather*}
\left\langle \sqrt{\rho_{1}},\sqrt{\rho_{2}}\right\rangle _{HS}=\left\langle
\sqrt{P_{1}\rho_{1}P_{1}},\sqrt{P_{2}\rho_{2}P_{2}}\right\rangle _{HS}\\
=\left\langle P_{1}\sqrt{\rho_{1}}P_{1},P_{2}\sqrt{\rho_{2}}P_{2}\right\rangle
_{HS}=\Tr\left(  P_{1}\sqrt{\rho_{1}}P_{1}P_{2}\sqrt{\rho_{2}}P_{2}\right)  =0.
\end{gather*}
Thus the usual treatment with sharp apparatus observables \cite{BS}\ amounts
to imposing
\[
\left\langle \sqrt{\rho_{1}},\sqrt{\rho_{2}}\right\rangle _{HS}=0
\]
for all $\rho_{1}\in A_{1}$, and for all $\rho_{2}\in A_{2}$.
\end{remark}

\begin{remark}
In \cite{B} unsharp apparatus observables (of a restricted type to allow for
definite apparatus configurations) are considered. In the present case of
distinguishing two apparatus configurations only, this amounts to introducing
a positive operator valued measure on the measure space $\left\{  1,2\right\}
$. Thus the projections $P_{i}$ are replaced by effects, i.e. linear
continuous operators $E_{i}:\mathcal{H}\rightarrow\mathcal{H}$ with $0\leq
E_{i}\leq id_{\mathcal{H}}$ and $E_{1}+E_{2}=id_{\mathcal{H}}$. The effects
$E_{i}$ are assumed to have the eigenvalue $1$ and a state $\rho$ with
definite apparatus position $i$ is assumed to obey $\Tr(E_{i}\rho)=1$.
Introducing an orthonormal basis of eigenvectors $e_{\alpha}$ of $\rho$ one
obtains the spectral representation
\[
\rho=\sum_{\alpha\in I}\lambda_{\alpha}P_{e_{\alpha}}%
\]
with $\sum_{\alpha\in I}\lambda_{\alpha}=1$\thinspace and $\lambda_{\alpha}>0$
for all $\alpha\in I$. Now $0\leq E_{i}\leq id_{\mathcal{H}}$ implies
$0\leq\left\langle e_{\alpha},E_{i}e_{\alpha}\right\rangle \leq1$ and from%
\[
1=\Tr(E_{i}\rho)=\sum_{\alpha\in I}\lambda_{\alpha}\left\langle e_{\alpha
},E_{i}e_{\alpha}\right\rangle
\]
one infers $\left\langle e_{\alpha},E_{i}e_{\alpha}\right\rangle =1$ for all
$\alpha\in I$. Therefore $E_{i}e_{\alpha}=e_{\alpha}$, i.e. every eigenvector
of $\rho$ with nonzero eigenvalue is an eigenvector of $E_{i}$ with eigenvalue
$1$. Let now $\rho_{1}$ and $\rho_{2}$ obey $\Tr(\rho_{i}E_{i})=1\ $and let $x$
and $y$ be eigenvectors of $\rho_{1}$ and $\rho_{2}$ for nonzero eigenvalues
respectively. Thus $E_{1}x=x$ and $E_{2}y=y$ follows. From this and
$E_{1}+E_{2}=id_{\mathcal{H}}$\ we obtain%
\[
\left\langle x,y\right\rangle =\left\langle x,(E_{1}+E_{2})y\right\rangle
=\left\langle x,E_{1}y\right\rangle +\left\langle x,E_{2}y\right\rangle
=\left\langle E_{1}x,y\right\rangle +\left\langle x,E_{2}y\right\rangle
=2\left\langle x,y\right\rangle .
\]
Thus $\left\langle x,y\right\rangle =0$ follows. From this we obtain
$\sqrt{\rho_{2}}x=0$ and from the spectral representation $\rho_{1}%
=\sum_{\alpha\in I}\lambda_{\alpha}P_{e_{\alpha}}$ finally%
\[
\left\langle \sqrt{\rho_{1}},\sqrt{\rho_{2}}\right\rangle _{HS}=\sum
_{\alpha\in I}\sqrt{\lambda_{\alpha}}\left\langle e_{\alpha},\sqrt{\rho_{2}%
}e_{\alpha}\right\rangle =0
\]
follows. Thus also Busch's treatment \cite{B} with unsharp apparatus
observables amounts to imposing $\left\langle \sqrt{\rho_{1}},\sqrt{\rho_{2}%
}\right\rangle _{HS}=0$ for all $\rho_{1}\in A_{1}$, and for all $\rho_{2}\in
A_{2}$.
\end{remark}

\begin{itemize}
\item Assumption 3: The time evolution under the measurement interaction from
a premeasurement instant of time to a postmeasurement one is assumed to be
given by some unitary operator $U:\mathcal{H}\mathcal{\rightarrow}\mathcal{H}%
$. The evolution of density operators is then the mapping $u:\mathcal{C}%
_{HS}(\mathcal{H})\rightarrow\mathcal{C}_{HS}(\mathcal{H}),a\mapsto
UaU^{\ast}$.
\end{itemize}

\begin{remark}
The linear mapping $u$ is a unitary algebra automorphism, i.e.
\[
u(ab)=u(a)u(b)\quad\text{and }\left\langle u\left(  a\right)  ,u\left(
b\right)  \right\rangle _{HS}=\left\langle a,b\right\rangle _{HS}.
\]
Thus for a density operator $\rho$ there holds $\sqrt{u(\rho)}=u(\sqrt{\rho})$
and furthermore
\[
\left\langle \sqrt{u\left(  \rho_{1}\right)  },\sqrt{u\left(  \rho_{2}\right)
}\right\rangle _{HS}=\left\langle u\left(  \sqrt{\rho_{1}}\right)  ,u\left(
\sqrt{\rho_{2}}\right)  \right\rangle _{HS}=\left\langle \sqrt{\rho_{1}}%
,\sqrt{\rho_{2}}\right\rangle _{HS}.
\]
\end{remark}

\begin{proposition}
Let the density operator $E$ from $\mathcal{H}_{E}$ be such that $\rho
_{s}\otimes E\in A_{1}$ for all density operators $\rho_{s}$ on $\mathcal{H}%
_{s}$. Assume $u\left(  P_{\psi_{i}}\otimes E\right)  \in A_{i}$ for $i=1,2$.
Then the pure state $P_{\phi}$ with $\phi:=\frac{1}{\sqrt{2}}(\psi_{1}%
+\psi_{2})\ $has a neighborhood $\mathcal{D}$ in the set of density operators
on $\mathcal{H}_{s}$ such that for all $\rho_{s}\in\mathcal{D}$ holds
$u\left(  \rho_{s}\otimes E\right)  \notin A_{1}\cup A_{2}$.
\end{proposition}

\begin{remark}
The proposition demonstrates that if the apparatus (irrespective of the
microsystem's state) is in configuration $1$ before the measurement and if for
the initial states $P_{\psi_{i}}\otimes E$ the unitary measurement dynamics
$u$ results in states with apparatus configuration $i$ then for an initial
state $\rho_{s}\otimes E$ with $\rho_{s}$ sufficiently close to $P_{\phi
}\otimes E$ the postmeasurement density operator $u\left(  \rho_{s}\otimes
E\right)  $ does not possess a definite postmeasurement apparatus configuration.
\end{remark}

\begin{proof}
Due to remark 5 we have%
\begin{align*}
\left\langle \sqrt{u\left(  P_{\phi}\otimes E\right)  },\sqrt{u\left(
P_{\psi_{i}}\otimes E\right)  }\right\rangle _{HS}  &  =\left\langle
\sqrt{P_{\phi}}\otimes\sqrt{E},\sqrt{P_{\psi_{i}}}\otimes\sqrt{E}\right\rangle
_{HS}\\
&  =\left\langle \sqrt{P_{\phi}},\sqrt{P_{\psi_{i}}}\right\rangle
_{HS}\left\langle \sqrt{E},\sqrt{E}\right\rangle _{HS}\\
&  =\left\langle P_{\phi},P_{\psi_{i}}\right\rangle _{HS}=\left|  \left\langle
\phi,\psi_{i}\right\rangle \right|  ^{2}=\frac{1}{2}.
\end{align*}
Thus $\left|  \left\langle \sqrt{u\left(  P_{\phi}\otimes E\right)  }%
,\sqrt{u\left(  P_{\psi_{i}}\otimes E\right)  }\right\rangle _{HS}\right|
>\varepsilon$. Since $u\left(  P_{\psi_{i}}\otimes E\right)  \in A_{i}$, the
vector $u\left(  P_{\phi}\otimes E\right)  $ neither belongs to $A_{1}$ nor to
$A_{2}$. Because of the continuity of $\left\langle \cdot,\cdot\right\rangle
_{HS}$, for all $\rho_{s}$ in a neighborhood $\mathcal{D}$ of $P_{\phi}$ also
$u\left(  \rho_{s}\otimes E\right)  \notin A_{1}\cup A_{2}$ holds.
\end{proof}

\section{Conclusion}

The QMP shows that so far standard quantum theory has no rule of how to
ascribe internal properties to general states of closed systems (comprising
observers) such that our definite everyday sensations find an explanation.
Standard quantum theory seems to need a splitting of the world into a quantum
part and a dynamically unresolved environment from which the quantum part is
observed and in relation to which properties can be induced by observation.
Depending on the way how the environment observes, such external observation
forces the quantum part into assuming properties by means of stochastic
quantum jumps. These jumps either lead into the specific set of states which
have the property under consideration or into the set of states which
definitely do not have this property. (Since the union of these two subsets is
unequal to the set of all states, complementarity of properties emerges.) All
this seems to tell that standard quantum theory cannot constitute a consistent
theoretical framework for describing arbitrarily large systems.

Bohmian mechanics enriches the conceptual framework of standard quantum
mechanics by degrees of freedom (''hidden variables''), such that with the
states of closed systems definite properties can be associated without making
reference to any external agent. Quantum jumps and state reduction do not
occur. States and their properties vary continuously and deterministically
with time. The QMP completely disappears.\cite{BDDGZ} If the Bohmian program
could be extended to the realm of relativistic quantum fields, a quantum frame
work with the potential for universal validity were found.


\begin{thebibliography}{9}                                                                                                %

\bibitem {BG}A. Bassi, G.C. Ghirardi, Phys. Lett. A 275 (2000) 373.

\bibitem {Neu}J. von Neumann, Mathematische Grundlagen der Quantenmechanik,
Springer, Berlin, 1932

\bibitem {Wig}E.P. Wigner, Am. Journ. Phys. 31 (1963) 6

\bibitem {dEs}B. d'Espagnat, Conceptual foundations of quantum mechanics,
Benjamin, Menlo Park, 1971

\bibitem {Fin}A. Fine, Phys. Rev. D 2 (1970) 2783

\bibitem {Shi}A. Shimony, Phys. Rev. D 9 (1974) 2321

\bibitem {BS}P. Busch, A. Shimony, Stud. Hist. Phil. Mod. Phys. 27 (1996) 397

\bibitem {B}P. Busch, Int. Journ. Theor. Phys. 37 (1998) 241

\bibitem {Bell}J.S. Bell, Physics world 3(8) (1990) 33

\bibitem {Adl}S.L. Adler, \textit{Why decoherence has not solved the
measurement problem: a response to P.W.Anderson}, arXiv: quant-ph/0112095

\bibitem {BDDGZ}K. Berndl, et al., Nuovo Cimento B 110 (1995) 737

\bibitem {DGZ}D. D\"{u}rr, S. Goldstein, N. Zanghi, \textit{Bohmian mechanics
as the foundation of quantum mechanics}, in: J.T. Cushing, A. Fine, S.
Goldstein, (Eds.) Bohmian mechanics and quantum theory: an appraisal, Kluwer,
Dordrecht, 1996

\bibitem {Due}D. D\"{u}rr, Bohmsche Mechanik als Grundlage der
Quantenmechanik, Springer, Berlin, 2001

\bibitem {Adle}S.L. Adler, Journ. Phys. A 35 (2002) 841

\bibitem {RS}M. Reed, B. Simon, Methods of modern mathematical physics, Vol.
1, Academic, New York, 1972 (sect. VI.6)
\end{thebibliography}
\end{document}